\documentstyle[NATO]{crckapb} 

\def\etal{{\it et al. }\rm}
\def\simlt{\mathrel{\hbox{\rlap{\hbox{\lower4pt\hbox{$\sim$}}}\hbox{$<$}}}}
\def\simgt{\mathrel{\hbox{\rlap{\hbox{\lower4pt\hbox{$\sim$}}}\hbox{$>$}}}}
\newcounter{parentequation}
\def\beglet{
  \addtocounter{equation}{1}%
  \setcounter{parentequation}{\value{equation}}%
  \setcounter{equation}{0}%
  \def\theequation{\arabic{parentequation}\alph{equation}}%
  \ignorespaces
}
\def\endlet{
  \setcounter{equation}{\value{parentequation}}%
  \def\theequation{\arabic{equation}}%
}

\begin{opening}
\title{CMB ANISOTROPIES AND THE DETERMINATION\protect\\
       OF COSMOLOGICAL PARAMETERS}

\subtitle{ }

\author{G EFSTATHIOU}
\institute{Institute of Astronomy\\
           Madingley Road\\
           Cambridge CB3 OHA}

\end{opening}

\runningtitle{CMB Anisotropies}

\begin{document}

\section{Abstract}
I review the basic theory of the cosmic microwave
background (CMB) anisotropies in adiabatic cold dark matter (CDM)
cosmologies. The latest observational results on the CMB power
spectrum are consistent with the simplest inflationary models and
indicate that the Universe is close to spatially flat with a nearly
scale invariant fluctuation spectrum. We are also beginning to see
interesting constraints on the density of CDM, with a best fit value
of $\omega_c \equiv \Omega_c h^2 \sim 0.1$. The CMB constraints, when
combined with observations of distant Type Ia supernovae, are
converging on a $\Lambda$-dominated Universe with $\Omega_m \approx
0.3$ and $\Omega_\Lambda \approx 0.7$.\footnote{To appear in Proceedings of NATO
ASI: Structure formation in the Universe, eds. N. Turok, R. Crittenden.}

\section{Introduction}

The discovery of temperature anisotropies in the CMB by the COBE team
(Smoot \etal 1992) heralded a new era in cosmology.  For the first
time COBE provided a clear detection of the primordial fluctuations
responsible for the formation of structure in the Universe at a time
when they were still in the linear growth regime. Since then, a large
number of ground based and balloon borne experiments have been
performed which have succeeded in defining the shape of the power
spectrum of temperature anisotropies $C_\ell$\footnote{The power
spectrum is defined as $C_\ell = \langle \vert a_{\ell m}\vert^2
\rangle$, where the $a_{\ell m}$ are determined from a spherical
harmonic expansion of the temperature anisotropies on the sky, $\Delta
T/T = \sum a_{\ell m} Y_{\ell m}(\theta, \phi)$.}  up to multipoles of
$\ell \sim 300$ clearly defining the first acoustic peak in the
spectrum. Figure 1 shows a compilation of band power anisotropy
measurements
\begin{equation}
 {\Delta T_\ell \over T} = \sqrt{ {1 \over 2 \pi} \ell (\ell + 1) C_\ell}
\end{equation}
that is almost up to date at the time of writing. The horizontal error
bars show the multipole range probed by each experiment. The recent
results from the VIPER experiment (Peterson \etal 1999) and the
Boomerang test flight (Mauskopf \etal 1999) are not plotted because
the exact window functions are not yet publically available. Neither
are the published results from the Python V experiment (Coble \etal
1999) which seem to be  discrepant with the other experiments particularly in
the multipole range $\ell \simlt 100$. The points plotted in figure 1
are generally consistent with each other and provide strong evidence
for a peak in the power spectrum at $\ell \sim 200$.

\begin{figure}
\vspace{9cm}  
\includegraphics{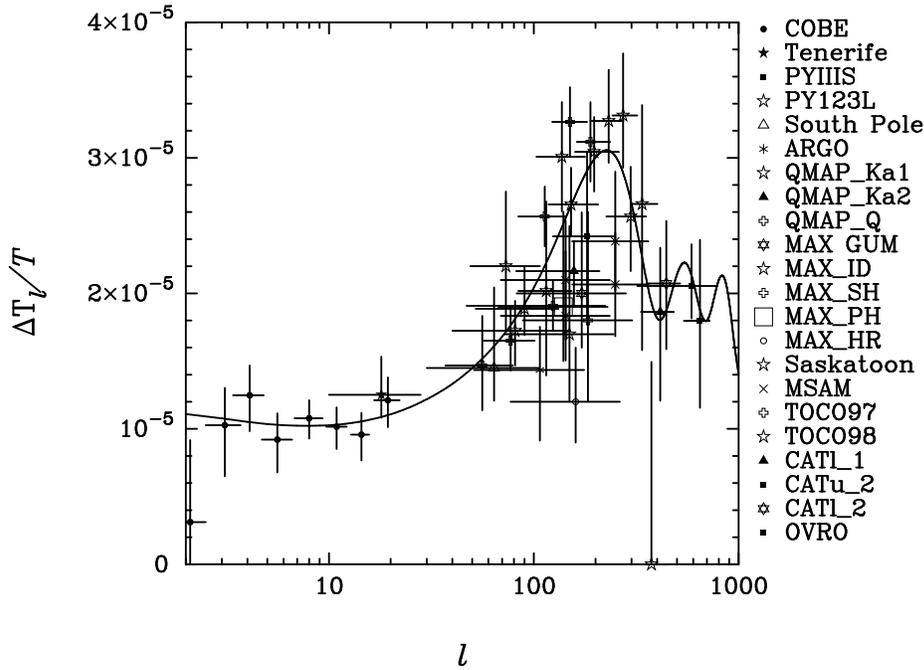}
\caption{Current constraints on the power spectrum of CMB temperature
anisotropies. The error bars in the vertical direction show
$1\sigma$ errors in the band power estimates and the error bars
in the horizontal direction indicate the width of the band.
The solid line shows the best fit adiabatic CDM model
with parameters $\omega_b = 0.019$, $\omega_c = 0.10$, $n_s = 1.08$,
$Q_{10} = 0.98$, $\Omega_m=0.225$, $\Omega_\Lambda = 0.775$.}
\end{figure}

In this introductory article, I will review briefly the theory of CMB
anisotropies in adiabatic models  of structure formation and then
discuss the implications of Figure 1 for  values of cosmological
parameters. The literature on the CMB anisotropies has grown
enormously over the last few years and it is impossible to do the
subject justice in a short article. General reviews of the CMB
anisotropies are given by Bond (1996) and Kamionkowski and Kosowsky
(1999). A recent review on constraining cosmological parameters from
the CMB is given by Rocha (1999).

\section{Basic Theory}

Most of the key features of figure 1 can be understood using a simplified
set of equations. The background universe is assumed to be spatially flat
together with small perturbations $h_{ij}$ so that the metric is 
\begin{eqnarray}
ds^2 = a^2(\tau)\left (\eta_{ij} + h_{ij} \right ) dx^i dx^j, \\
\qquad \qquad \eta_{ij} = (1,\ -1,\ -1,\ -1), \qquad \tau = \int dt/a. \nonumber
\end{eqnarray}
We adopt the synchronous gauge, $h_{00} = h_{0i} = 0$, and ignore the
anisotropy of Thomson scattering and perturbations in the relativistic
neutrino component. With these assumptions, the equations governing
the evolution of scalar plane wave perturbations of wavenumber $k$ are
\beglet
\begin{eqnarray}
\dot \Delta + ik\mu \Delta +\Phi = \sigma_T n_e a \left
[\Delta_0 + 4 \mu v_b - \Delta \right ] \\
\Phi = - (3 \mu^2 -1)\dot h_{33}-(1 - \mu^2)\dot h \\
\dot v_b + {\dot a \over a} v_b = \sigma_T n_e a {\bar \rho_\gamma \over \bar 
\rho_b} \left
( \Delta_1 - {4 \over 3} v_b \right ),  \\
\dot \delta_b = {1 \over 2} \dot h - ikv_b, \qquad
\dot \delta_C = {1 \over 2} \dot h \qquad\\
\ddot h + {\dot a \over a}\dot h = 8\pi G a^2 (\bar \rho_b \delta_b
+ \bar \rho_{c} \delta_{c} + 2 \bar \rho_\gamma \Delta_0) \\
ik(\dot h_{33} - \dot h) = 16 \pi G a^2 (\bar \rho_b v_b + \bar
\rho_\gamma \Delta_1).
\end{eqnarray}
\endlet 
Here, $\Delta$ is the perturbation to the photon radiation brightness
and $\Delta_0$ and $\Delta_1$ are its zeroth and
first angular moments, $\delta_b$ and $\delta_c$ are the density
perturbations in the baryonic and CDM components, $v_b$ is the baryon
velocity and $h = {\rm Tr}(h_{ij})$. Dots denote differentiation with
respect to the conformal time variable $\tau$. It is instructive to
look at the solutions to these equations in the limits of large ($k
\tau_R \ll 1$) and small ($k \tau_R \gg 1$) perturbations, where
$\tau_R$ is the conformal time at recombination:

\subsection{Large angle anisotropies}

 In the limit ${k} \tau_R \ll 1$, Thomson scattering is unimportant and so the term in square
brackets in the Boltzmann equation for the photons  can be ignored. In the
matter dominated era $h_{33} = h \propto \tau^2$ and so equation (3a) becomes
\begin{equation}
\dot \Delta + ik\mu\Delta = 2\mu^2\dot h \label{LA1}
\end{equation}
with  approximate solution 
\begin{equation}
\Delta (k,\mu,\tau) \approx  - {2 \ddot h (\tau_R) \over k^2} \exp \left
(ik\mu (\tau_s - \tau) \right ). \label{LA2}
\end{equation}
This solution is the Sachs-Wolfe (1967) effect. Any deviation from the
evolution $\ddot h ={\rm constant}$, caused for example by a non-zero
cosmological constant, will lead to additional terms in equation
(\ref{LA2}) increasing the large-angle anisotropies (sometimes
referred to as the late-time Sachs-Wolfe effect, see {\it e.g.} Bond
1996).  The CMB power spectrum is given by
\begin{equation}
C_\ell = {1 \over 8\pi} \int^\infty_0 \vert \triangle_\ell \vert^2
k^2dk, \label{LA3}
\end{equation} 
where the perturbation $\Delta$ has been expanded in Legendre 
polynomials,
\begin{equation}
\Delta = {\displaystyle \sum_{\ell}} (2\ell+1)\Delta_\ell P_\ell (\mu). \label{LA4}
\end{equation}
Inserting the solution of equation (\ref{LA2}) into  equation (\ref{LA3}) gives
\begin{equation}
C_\ell =  {1 \over 2 \pi}\int^\infty_0 {\vert \ddot h \vert^2 \over
k^4} j_l^2 (k \tau_0)k^2\; dk, \label{LA5}
\end{equation}
and so for a power-law spectrum of scalar perturbations, $\vert h \vert^2 \propto k^{n_s}$,
the CMB power spectrum is
\begin{equation}
C_\ell = C_2 { \Gamma \left ( \ell + {(n_s - 1) \over 2} \right ) \over \Gamma 
\left ( \ell + {(5 - n_s) \over 2} \right ) } { \Gamma \left ( { 9 - n_s \over 2} \right ) 
\over \Gamma \left ( {3 + n_s \over 2} \right ) } 
\end{equation}
giving the characteristic power-law like form, $C_\ell \propto \ell^{(n_s -3)}$ at low
multipoles ($\ell \simlt 30$).

\subsection{Small angle anisotropies and Acoustic peaks}

In the matter dominated era, equation (3a) becomes
\begin{equation}
\dot\Delta + ik\mu \triangle = \sigma_Tn_e a \left [ \Delta_0
+ 4\mu v_b -\Delta \right ] + 2\mu^2\dot h, \label{SA1}
\end{equation}
and taking the zeroth and first angular moments gives
\beglet
\begin{eqnarray}
\dot\Delta_0 + ik \Delta_1 = {2 \over 3} \dot h \qquad \qquad\\
\dot\Delta_1 + ik \left ({\Delta_0 + 2 \Delta_2 \over 3 }\right ) = 
\sigma_Tn_e a \left [{4 \over 3} v_b  - \Delta_1 \right ].
\end{eqnarray}
\endlet
Prior to recombination, $\tau/\tau_c \gg 1$
where $\tau_c = 1/(\sigma_T n_e a)$ is the mean collision time,  and so
the matter is tightly coupled to the radiation. In this limit
 $\Delta_1 \approx 4/3v_b$ from equation (3c) and 
$\Delta_2$ in equation (11b) can be ignored. With these approximations, equation (11b)
becomes
\begin{eqnarray}
\dot\Delta_1 + {ik\Delta_0 \over 3} = -{\bar \rho_b \over \bar 
\rho_\gamma}\left [{3 \over 4}
 \dot\Delta_1 + {\dot a \over a} \Delta_1 \right ].
\end{eqnarray}
Neglecting the expansion of the universe, equations (11a) and (12) can be combined to
give a forced oscillator equation
\begin{eqnarray}
\ddot\Delta_0 = -{k^2 \over 3R} \Delta_0 + {2 \over 3}\ddot h, \qquad \ \ \
R \equiv 1+{3 \bar \rho_b \over 4 \bar \rho_\gamma },
\end{eqnarray}
with solution
\begin{eqnarray}
\Delta_0(\tau)=\left (\Delta_0 (0) - {2R\ddot h \over k^2}\right ) \cos {k\tau \over \sqrt {3R}} + {\sqrt {3R} \over k} \dot\Delta_0 (0)\sin {k\tau \over \sqrt {3R}} + {2R \ddot h \over k^2},
\end{eqnarray}
where $\Delta_0(0)$ and $\dot \Delta_0(0)$ are evaluated when the wave first crosses the Hubble
radius, $k \tau \sim 1$. For adiabatic perturbations the first term dominates over the second
because the perturbation breaks at $k \tau \sim 1$ with $\dot \Delta_0 \approx 0$. It is useful
to define (gauge-invariant) radiation perturbation variables
\begin{eqnarray}
\tilde\Delta_0 =  \Delta_0 - { 2\ddot h \over k^2},  \qquad
\tilde\Delta_1 =  \Delta_1+i{ 2 \dot h \over 3k}, \nonumber 
\end{eqnarray}
then the solution of equation (10) is
\begin{eqnarray}
\tilde\Delta (k, \mu, \tau ) = \int^\tau_0 \sigma_Tn_ea \left [\tilde
\Delta_0 + 4\mu \left (v_b + { i\dot h \over 2 k} \right ) \right ]
e^{ik\mu(\tau^\prime-\tau)-\int_{\tau^\prime}^\tau [\sigma_T n_e
a]d\tau^{\prime\prime}}d\tau^\prime.  \label{SA2}
\end{eqnarray}
If $k \tau \gg 1$, the second term in the square brackets is smaller
than the first by a factor of $k\tau$, and the solution of equation
(\ref{SA2}) gives a power spectrum with a series of modulated acoustic
peaks spaced at regular intervals of $k_m r_s(a_r) = m \pi$, where
$r_s$ is the sound horizon at recombination
\begin{eqnarray}
r_s  = {c \over {\sqrt 3}H_0\Omega_m^{1/2}}\int^{a_r}_0 {
da \over \left (a + a_{equ}\right )^{1/2}} {1 \over R^{1/2}}, \label{SA3}
\end{eqnarray}
(Hu and Sugiyama 1995). Here $a_{equ}$ is the scale factor when matter
and radiation have equal densities and $a_r$ is the scale factor at
recombination.

The multipole locations of the acoustic peaks in the angular power spectrum  are given 
by
\begin{eqnarray}
{\ell}_m = \alpha m\pi {d_A (z_r)\over r_s}
\end{eqnarray}
where $\alpha$ is a number of order unity and 
$d_A$ is the angular diameter distance to last scattering 
\beglet
\begin{eqnarray}
d_A =  {c \over H_0\vert \Omega_k \vert ^{1/2}} {\rm sin}_k( \vert \Omega_k \vert 
^{1/2} x)\\
x \approx \int^1_{a_r} {da \over [\Omega_ma + \Omega_ka^2 + \Omega_\Lambda a^4]^{1/2}}
\end{eqnarray}
\endlet
where $\Omega_k =  1-\Omega_\Lambda - \Omega_m$ and ${\rm sin}_k \equiv {\rm sinh}$ if
$\Omega_k > 0$ and ${\rm sin}_k = {\rm sin}$ if $\Omega_k < 0$.

The general dependence of the CMB power spectrum on cosmological
parameters is therefore clear. The positions of the acoustic peaks
depend on the geometry of the Universe via the angular diameter
distance of equation (18) and on the value of the sound horizon
$r_s$. The relative amplitudes of the peaks depend on the physical
densities of the various constituents $\omega_b \equiv \Omega_b h^2$,
$\omega_c \equiv \Omega_c h^2$, $\omega_\nu \equiv \Omega_\nu h^2$,
{\it etc.} and on the scalar fluctuation spectrum (parameterized here
by a constant spectral index $n_s$). Clearly, models with the same
initial fluctuation spectra and identical physical matter densities
$\omega_i$ will have identical CMB power spectra at high multipoles if
they have the same angular diameter distance to the last scattering
surface. This leads to a strong {\it geometrical degeneracy} between
$\Omega_m$ and $\Omega_\Lambda$ ({\it e.g.} Efstathiou and Bond 1999
and references therein). The power spectrum on large angular scales
(equation 9) is sensitive to the spectral index and amplitude of the
power spectrum, geometry of the Universe and, for extreme values of
$\Omega_k$ can break the geometrical degeneracy via the late-time
Sachs-Wolfe effect.  We will discuss briefly some of the constraints
on cosmological parameters from the current CMB data in the next
section. Before moving on to this topic, I mention some important
points that cannot be covered in detail because of space limitations:

\noindent
$\bullet$ Inflationary models can give rise to tensor perturbations
with a characteristic spectrum that declines sharply at $\ell \simgt
100$ (see {\it e.g.} Bond 1996 and references therein). In power-law
like inflation, the tensor spectral index $n_t$ is closely linked to
the scalar spectral index, $n_t \approx n_s -1$, and to the relative
amplitude of the tensor and scalar perturbations.

\noindent
$\bullet$ The anisotropy of Thomson scattering causes the CMB
anisotropies to be linearly polarized at the level of a few
percent (see Bond 1996, Hu and White 1997, and references
therein). Measurements of the linear polarization can distinguish
between tensor and scalar perturbations and can constrain the epoch of
reionization of the intergalactic medium (Zaldarriaga, Spergel and
Seljak 1997).

\noindent
$\bullet$ The main effect of reionization is to depress the amplitude
of the power spectrum at high multipoles by a factor of ${\rm
exp}(-2\tau_{opt})$ where $\tau_{opt}$ is the optical depth to Thomson
scattering. In the `best fit' CDM universe described in the next
section ($\omega_b = 0.019$, $h = 0.65$, $\Omega_m = \Omega_c +
\Omega_b \approx 0.3$ and $\Omega_\Lambda \approx 0.7$) and a
reionization redshift of $z_{reion} \approx 20$ (a plausible value)
$\tau_{opt} \approx 0.2$ which is significant. There is a reasonable chance
that we might learn something about the `dark ages' of cosmic history from
precision measurements of the CMB.

\begin{figure}
\vspace{5.0cm} 
\includegraphics{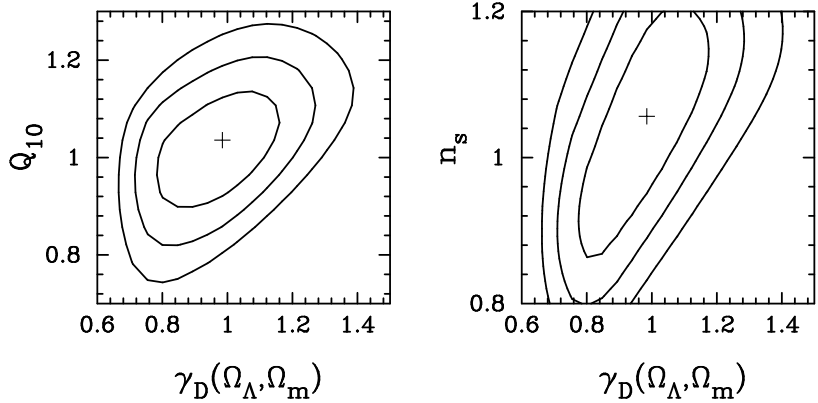}
\caption{Marginalized likelihoods ($1$, $2$ and $3\sigma$ contours) in
the $Q_{10}$--$\gamma_D$ and $n_s$--$\gamma_D$ planes, where
$\gamma_D$ is the acoustic peak location parameter defined in
equation 18.}

\end{figure}

\begin{figure}
\vspace{5.0cm} 
\includegraphics{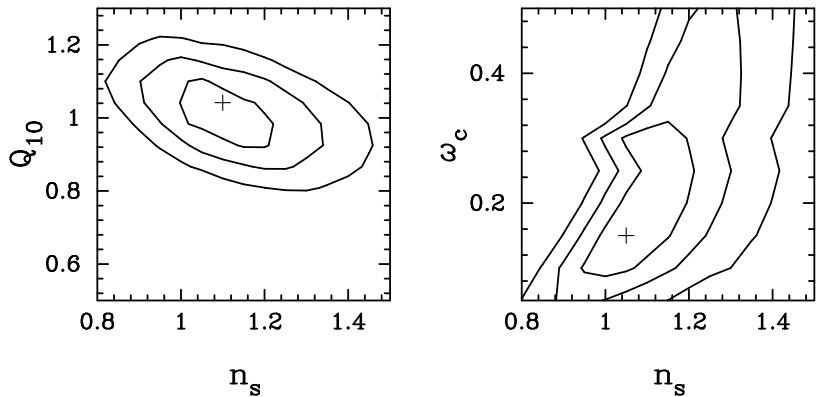}
\caption{Marginalized likelihoods ($1$, $2$ and $3\sigma$ contours) in
the $Q_{10}$--$n_s$ and $\omega_c$--$n_s$ planes. The crosses show where the
likelihood function peaks.}
\end{figure}

\section{Cosmological Parameters from the CMB}

In this section, we review some of the constraints on cosmological
parameters from the CMB data plotted in figure 1. The analysis is
similar to that presented in Efstathiou \etal (1999, hereafter E99),
in which we map the full likelihood function in $5$ parameters
$\Omega_\Lambda$, $\Omega_m$, $\omega_c$, $n_s$ and $Q_{10}$ (the
amplitude of $\sqrt{C_\ell}$ at $\ell = 10$ relative to that inferred
from COBE). The baryon density is constrained to $\omega_b = 0.019$,
as determined from primordial nucleosynthesis and deuterium abundances
measurements from quasar spectra (Burles and Tytler 1998). The results
presented below are insensitive to modest variations ($\sim 25 \%$) of
$\omega_b$ and illustrate the main features of cosmological parameter
estimation from the CMB. Recently, Tegmark and Zaldarriaga (2000) have
performed a heroic $10$ parameter fit to the CMB data, including a
tensor contribution and finite optical depth from reionization. I will
discuss the effects of widening the parameter space briefly below but
refer the reader to Tegmark and Zaldarriaga for a detailed analysis.

The best fit model in this five parameter space is plotted as the
solid line in figure 1. It is encouraging that the best fitting model
has perfectly reasonable parameters, a spatially flat universe with a
nearly scale invariant fluctuation spectrum and a low CDM density
$\omega_c \sim 0.1$. Marginalised likelihood functions are plotted
in various projections in the parameter space in figures 2, 3 and 5
(uniform priors are assumed in computing the marginalized likelihoods,
as described in E99). Figure 2 shows constraints on the position of
the first acoustic peak measured by the `location' parameter
\begin{eqnarray}
\gamma_D = {\ell_D (\Omega_\Lambda, \Omega_m) \over
\ell_D(\Omega_\Lambda = 0, \Omega_m = 1)},
\end{eqnarray}
{\it i.e.} the parameter $\gamma_D$ measures the location of the
acoustic peak relative to that in a spatially flat model with zero
cosmological constant. The geometrical degeneracy between $\Omega_m$
and $\Omega_\Lambda$ described in the previous section is expressed by
$\gamma_D = {\rm constant}$. Figure 2 shows that the best fitting
value is $\gamma_D = 1$ with a $2\sigma$ range of about $\pm 0.3$. The
position of the first acoustic peak in the CMB data thus provides
powerful evidence that the Universe is close to spatially flat.

\begin{figure}
\vspace{7cm} 
\includegraphics{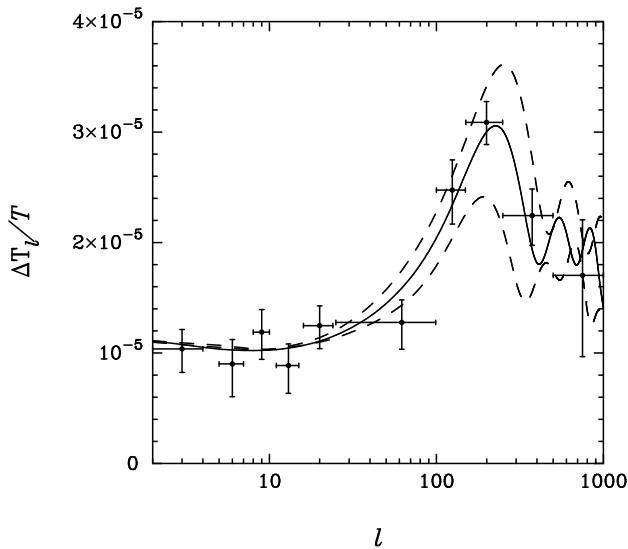}
\caption{The crosses show maximum likelihood bandpower averages of the
observations shown in figure 1 together with  $1\sigma$ errors. The solid line
shows the best fit adiabatic CDM model as plotted in figure 1 which
has $\omega_c = 0.1$. The dashed lines show the effects of varying
$\omega_c$ keeping the other parameters fixed. The upper dotted line
shows $\omega_c = 0.05$ and the lower dashed line shows $\omega_c = 0.25$.}
\end{figure}

Figure $3$ shows the marginalized likelihoods in the $Q_{10}- n_s$ and
$\omega_c - n_s$ planes. The constraints on $Q_{10}$ and $n_s$ are not
very different to those from the analysis of COBE alone (see {\it
e.g.} Bond 1996). The experiments at higher multipoles are so
degenerate with variations in other cosmological parameters that they
do not help tighten the constraints on $Q_{10}$ and $n_s$. The
constraints on $\omega_c$ and $n_s$ show an interesting result; if
$n_s \approx 1$, then the best fit value of $\omega_c$ is about $0.1$
with a $2\sigma$ upper limit of about $0.3$. This constraint on
$\omega_c$ comes from the height of the first acoustic peak, as shown
in figure 4. In this diagram, the CMB data points have been averaged
in $10$ band-power estimates as described by Bond, Knox and Jaffe
(1998).  The solid curve shows the best-fit model as plotted in figure
1, which has $\omega_c = 0.1$. The dashed lines show models with
$\omega_c = 0.25$ and $\omega_c = 0.05$ with the other parameters held
fixed. Raising $\omega_c$ lowers the height of the peak and
vice-versa. This result is not very sensitive to variations of
$\omega_b$ in the neighbourhood of $\omega_b \sim 0.02$. Reionization
and the addition of a tensor component can lower the height of
the first peak relative to the anisotropies at lower multipoles
and so the upper limits on $\omega_c$ are robust to the addition
of these parameters. The CMB data have now reached the point where
we have good constraints on the height of the first peak, as well
as its location,  and this is beginning to set interesting constraints
on $\omega_c$. The best fit value of  $\Omega_m \approx 0.3$, derived
from combining the CMB data with results from distant Type Ia
supernovae (see figure 5) implies $\omega_c \approx 0.11$ for a Hubble constant
of $h = 0.65$, consistent with the low values of $\omega_c$ favoured by the
height of the first acoustic peak. 

The left hand panel of figure 5 shows the marginalized likelihood for
the CMB data in the $\Omega_\Lambda$--$\Omega_m$ plane. The likelihood
peaks along the line for spatially flat universes $\Omega_k=0$ and it
is interesting to compare with the equivalent figure in E99 to see how
the new experimental results of the last year have caused the likelihood
contours to narrow down around $\Omega_k =0$. (See also Dodelson and
Knox 1999 for a similar analysis using the latest CMB data).  As is
well known, the magnitude-redshift relation for distant Type Ia
supernovae results in nearly orthogonal constraints in the
$\Omega_\Lambda$--$\Omega_m$ plane, so combining the supernovae and
CMB data can break the geometrical degeneracy. The right hand panel in
Figure 5 combines the CMB likelihood function derived here with the
likelihood function of the supernovae sample of Perlmutter \etal
(1999) as analysed in E99. The combined likelihood function is peaked
at $\Omega_m \approx 0.3$ and $\Omega_\Lambda \approx 0.7$. 

\begin{figure}
\vspace{6.5cm}  

\includegraphics{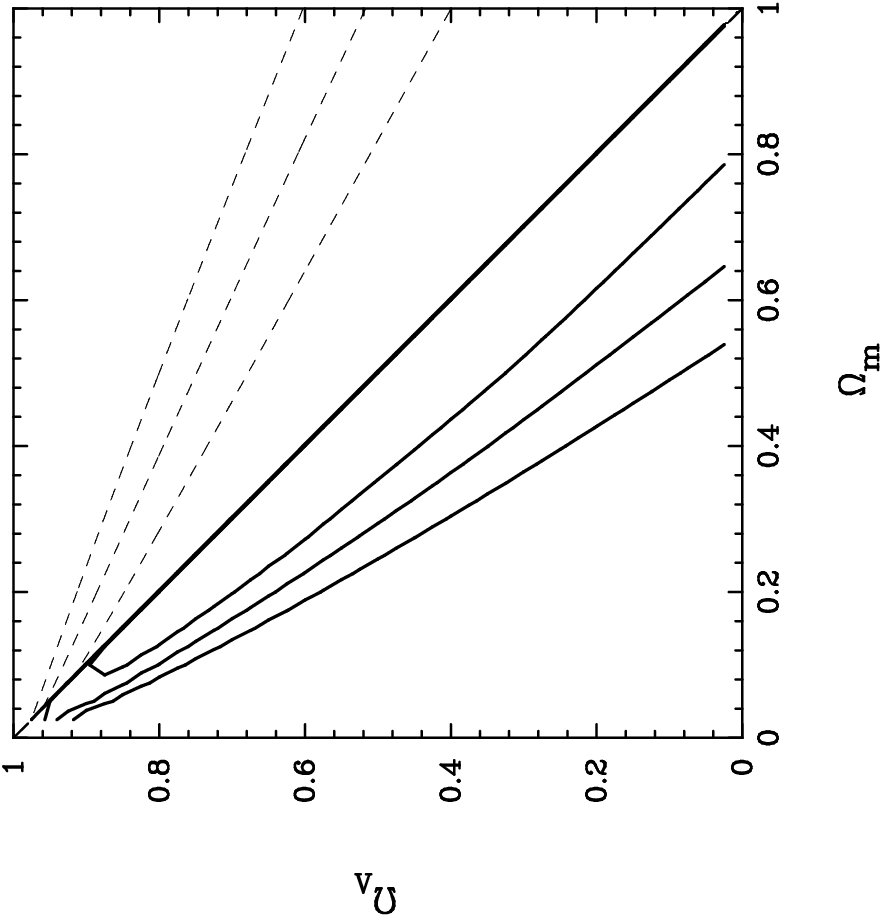}

\includegraphics{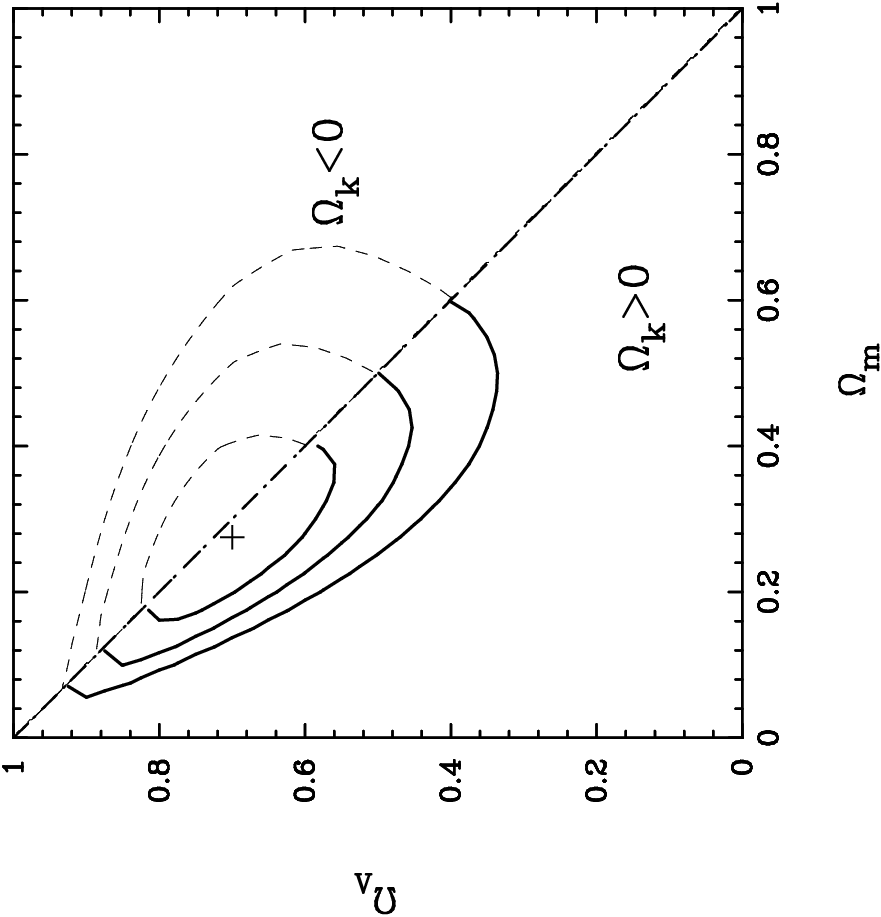}

\caption{The figure to the left shows the $1$, $2$ and $3\sigma$
likelihood contours marginalized in the $\Omega_\Lambda$ and
$\Omega_m$ plane from the observations plotted in figure 1. The figure
to the right shows the CMB likelihood combined with the likelihood
function for Type Ia supernovae of Permutter {\it et al.} (1999) as
analyzed by E99.  The dotted contours in both figures extend the
CMBFAST (Seljak and Zaldarriaga 1996)
computations into the closed universe domain using the
approximate method described in E99.}
\end{figure}

It is remarkable how the CMB data and the supernovae data are homing
in on a consistent set of cosmological parameters that are compatible
with the simplest inflationary models and also with parameters inferred
from a number of other observations ({\it e.g.} galaxy clustering, 
baryon content of clusters and dynamical estimates of the mean mass density,
see Bahcall \etal 1999 for a review). It is also remarkable
that the `best fit' model requires a non-zero cosmological constant, a result that
few cosmologists would have thought likely a few years ago.

The next few years will see a revolutionary increase in the volume and
quality of CMB data. The results of the Boomerang Antarctic flight are
awaited with great interest and should be of sufficient quality to
render all previous analyses of cosmological parameters from the CMB
obsolete. The polarization of the CMB has not yet been discovered, but
a number of ground based and balloon borne experiments designed to
detect polarization are under construction (Staggs, Gundersen and
Church 1999). The MAP satellite, scheduled for launch in late 2000, will
have polarization sensitivity and should determine the power spectrum
$C_\ell$ accurately to about $\ell \sim 800$,  defining the first three
acoustic peaks. Further into the future, the Planck satellite,
scheduled for launch in 2007, should determine the power spectrum to
$\ell \simgt 2500$, provide sensitive polarization measurements and
extremely accurate subtraction of foregrounds. Evidently, the era of
precision cosmology is upon us and the next decade should see a
dramatic improvement in our knowledge of fundamental cosmological
parameters and in our understanding of the origin of fluctuations in
the early Universe.

\end{document}